# Characteristics of Nickel Thin Film and Formation of Nickel Silicide by Remote Plasma Atomic Layer Deposition using Ni($^i$Pr–DAD)$_2$

**Jinho Kim and Woochool Jang**

*Division of Materials Science and Engineering, Hanyang University, Seoul 133-791, Korea*

**Jingyu Park, Heeyoung Jeon, and Hyunjung Kim**

*Department of Nano-scale Semiconductor Engineering, Hanyang University Seoul 133-791, Korea*

**Junhan Yuh**[*]

*Corporate Technology Division, POSCO, Seoul 135-777, Korea*

**Hyeongtag Jeon**[*]

*Department of Materials Science and Engineering, Hanyang University, Seoul 133-791, Korea*

*Department of Nano-scale Semiconductor Engineering, Hanyang University Seoul 133-791, Korea*

In this study, the characteristics of nickel thin film deposited by remote plasma atomic layer deposition (RPALD) on p–type Si substrate and formation of nickel silicide using rapid thermal annealing were determined. Bis(1,4-di-isopropyl-1,3-diazabutadienyl)nickel, Ni($^i$Pr–DAD)$_2$, was used as a Ni precursor and ammonia plasma was used as a reactant. This was the first attempt to deposit Ni thin film using Ni($^i$Pr-DAD)$_2$ as a precursor for the ALD process. The RPALD Ni film was deposited with a growth rate of around 2.2Å/cycle at 250 °C and showed significant low resistivity of 33 μΩ·cm with a total impurity concentration of around 10 at. %. The impurities of the thin film, carbon and nitrogen, were existent by the forms of C–C and C–N in a bonding state. The impurities removal tendency was investigated by comparing of experimental conditions, namely process temperature and pressure. Nitrogen impurity was removed by thermal desorption during each ALD cycle and carbon impurity was reduced by the optimizing of the process pressure which is directly related with a mean free path of NH$_3$ plasma. After Ni deposition, nickel silicide was formed by RTA in a vacuum ambient for 1 minute. A nickel silicide layer from ALD Ni and PVD Ni was compared at the annealing temperature from 500 to 900 °C. NiSi from ALD Ni showed better thermal stability due to the contribution of small amounts of carbon and nitrogen in the as-deposited Ni thin film. Degradation of the silicide layer was effectively suppressed with a use of ALD Ni.

*Corresponding author

E-mail: junhan.yuh@gmail.com

E-mail: hjeon@hanyang.ac.kr

# I. INTRODUCTION

As memory device feature size has been scaling down for several decades, the contribution of the contact resistance has been increasing much more in total device resistance. These days, contact resistance has become a critical issue in current device fabrication technology. Thus, to reduce the contact resistance in a ULSI (ultra large scale integrated–circuit) device, the metal silicide technique is used for sub-micrometer devices [1-3]. Titanium was first used to form a salicide (self–aligned silicide) due to the low resistivity of $TiSi_2$ (12–14 $\mu\Omega\cdot cm$) compared with other silicide materials. Then $TiSi_2$ was replaced with $CoSi_2$ because the resistance of $TiSi_2$ had rapidly increased below the 0.3 μm technology [4,5]. But $CoSi_2$ has also had an abrupt increase in resistance caused by void formation in narrow poly silicon gates under the 65 nm technology node. Furthermore, the $CoSi_2$ phase requires a high silicon consumption ratio (Co:Si = 1:3.6) which is a critical problem in forming an ultra–shallow junction (USJ) [6]. Therefore, new silicide material is needed that satisfies the low resistivity and low silicon consumption ratio. In this regard, NiSi are promising metal silicide materials for next generation devices. NiSi has a relatively low silicon consumption ratio (Ni : Si = 1 : 1.8), low silicide formation temperature, and less line width dependence [7]. In the formation of NiSi, Ni atoms diffuse into the Si region. Thus, there is no bridging failure that is caused by diffusion of Si atoms between gate and source/drain in MOSFET structures [7,8]. Because of the numerous advantages, many researchers have studied the deposition of thin and conformal nickel film.

In conventional metal deposition, the PVD (physical vapor deposition) method has been widely used due to its high film quality without impurities. However, as the semiconductor device has a complex structure that includes a three–dimensional structure, and thus step coverage has become a great issue in film deposition. Therefore the CVD (chemical vapor deposition) or ALD method, which has good step coverage, should be investigated for metal deposition. The ALD method in particular, is very advantageous because it can control the film thickness by monolayer. Thus the ALD metal film could be expected to have excellent step coverage in terms of high aspect ratio [9]. Above all, the ALD method has a lot of advantages including low impurity concentration, conformal film formation, and a relatively low deposition temperature compared with the CVD method.

Based on the advantages of the ALD method, in this study Ni thin film was deposited using a remote plasma atomic layer deposition (RPALD) and NiSi was formed by the RTA process. There are several precursors for Ni deposition, but the result is a lot of carbon impurities remaining after deposition according to previous reports [10-12]. Generally, the impurity in thin film causes increases of sheet resistance, so the reduction of the impurities was especially important for the contact materials. In this study, a relatively pure Ni thin film was deposited by the ALD method using a newly introduced precursor, Ni($^i$Pr-DAD)$_2$. Furthermore, impurities in their binding states in deposited Ni film have been described and the changes of impurity concentration were investigated as a function of the ALD process parameters. And the formation of NiSi was performed with the RTA process at various annealing temperatures. The NiSi films from ALD Ni and PVD Ni were compared regarding thermal stability and surface morphology by the several analyses.

## II. EXPERIMENT

A Boron doped p–type Si (100) wafer with a resistivity of 1–10 Ω·cm was used for Ni thin film deposition using RPALD method. Bis(1,4–di–isopropyl–1,3–diazabutadienyl)nickel, Ni($^i$Pr–DAD)$_2$, was used as a precursor for the first time for the ALD process, and NH$_3$ plasma was used as a reducing agent. The Si substrate was cleaned with a dilute HF solution (HF : H$_2$O = 1 : 50) for 2 minutes to remove native oxide. After the substrate cleaning, a Si wafer was immediately loaded into the load–lock chamber to prevent native oxide formation and transferred to the deposition chamber. The vapor pressure of the precursor was 0.25 Torr at 93 °C and the canister for the Ni precursor was heated to 100 °C to supply the precursor into a chamber. The gas delivery lines were maintained at 130 °C to prevent contamination. The Ni thin film was deposited at a substrate temperature range from 150 to 300 °C and the process pressure from 0.6 to 1.8 Torr, respectively. These efforts were conducted to investigate the impurity concentration changing patterns as a function of the deposition parameter. Next, 400W $NH_3$ plasma was induced by inductively coupled plasma (ICP) type 13.56 MHz radio frequency (RF) power source with an ultra–high purity ammonia gas (99.99999%). For the ALD process, an injected Ni precursor and reactant $NH_3$ plasma were separated by purge Ar gas. Consequently, one cycle of the ALD process had four steps including 2 seconds of precursor injection with carrier Ar 50 SCCM (Standard Cubic Centimeter per Minutes), 8 seconds of precursor purge (Ar 100 SCCM), 10 seconds of reactant exposure ($NH_3$ gas 50–400 SCCM), and 8 seconds of a remaining gas purge (Ar 100 SCCM).After the deposition, Ni films were annealed to form a silicide phase. To compare the properties of NiSi, Ni films were prepared by the ALD method at a deposition temperature of 250 °C and the PVD method using e-beam evaporation of around 30nm thickness. An annealing process was conducted by RTA for 1 minute in a vacuum ambient ($1 \times 10^{-3}$ Torr) at the temperature range from 500 to 900 °C.

The electric resistance of the deposited Ni thin film and nickel silicide film were measured using a four point probe (FPP). Field emission scanning electron microscopy (FE–SEM) analysis was conducted to observe the deposited Ni film thickness and surface morphology of NiSi. Atomic concentration and chemical binding states were analyzed by auger electron spectroscopy (AES) and x–ray photoelectron spectroscopy (XPS). The crystal structures of the Ni film and NiSi layer were examined by x–ray diffraction (XRD).

## III. RESULTS AND DISCUSSION

Thin and conformal Ni thin film was deposited on a p–type Si substrate by remote plasma atomic layer deposition. Figure 1 (a) shows the growth rate of RPALD Ni deposited with Ni($^i$Pr–DAD)$_2$ and 400W NH$_3$ plasma at various substrate temperatures. The growth rate was gradually increased as deposition temperatures increases, and the increment of growth rate was culminated with 2.2 Å/cycle at 250 °C. Then, growth rate was slightly decreased at the substrate temperature of 300 °C. The resistivity of the deposited Ni film was calculated based on the film thicknesses and sheet resistance. Figure 1 (b) represents the changes of the Ni film resistivity in accordance with deposition temperature after 200 cycle deposition. The Ni thin film showed high resistivity at a deposition temperature of 150 °C at about 125 μΩ·cm, but it decreased with increasing substrate temperature. The resistivity of the deposited film was abruptly lowered at 250 °C and corresponded to a significant lower 33μΩ·cm. Then, the resistivity was slightly increased above 250 °C. To investigate the relationship between film resistivity and impurity concentrations, AES analysis was performed under the changes of process temperature and pressure.

Figure 2 shows the plot of the changes of major impurities in the film, nitrogen and carbon, as a function of process temperatures and pressure. As shown in figure 2 (a), the nitrogen concentration was continuously decreased depending on the temperature increases until 250 °C. At the temperature above 250°C, a similar amount of nitrogen was contained in the thin film. It appears that the decrease of nitrogen concentration is related to thermal desorption of nitrogen atoms from the Ni thin film surface. Meanwhile, the changes in the carbon contents showed no consistent trend in all temperature ranges. Furthermore, the carbon impurity concentration showed a dependence on the process pressure as indicated in figure 2 (b). In the case of process pressure of 0.6 Torr, a lot of carbon impurities were contained in the Ni thin film at about 30 at.%. Meanwhile, the carbon concentration was rapidly decreased at 1.2 Torr and slightly increased over the pressure of 1.8 Torr. The nitrogen concentration was not affected by the changes of the process pressure. These carbon and nitrogen impurities binding states were analyzed by following an XPS study.

The elemental composition and chemical binding states of the deposited Ni thin film were analyzed by XPS. Before the measurement, Ar sputtering on the surface was performed to avoid the detection of surface oxide and contamination of the Ni surface. All collected spectra were arranged by carbon impurity at 284.5 eV due to the charging effect. As shown in Figure 3 (a), the C 1s spectra was deconvoluted into three subpeaks corresponding to C–C, C–O [13,14] and C–N [15,16] bonds. These binding states constitute the C1s spectra by the C–C bond of 71.4 %, the C–O bond of 9.3 %, and the C–N bond of 19.3 % according to the area ratio of the spectra. Also, figure 3 (b) show that the N–C bond was observed at 397.5 eV in the N 1s spectra [17,18]. Thus, it was confirmed that carbon and nitrogen impurities exist not only in C–C bonding but also in C–N bonding state. This existence of the C-N bond could describe the nitrogen removal process in the following investigation.

Based on the analyzed results and literature study, the changes of carbon and nitrogen concentration were directly compared by an AES depth profile. Figure 4 represent the AES depth profiles of the Ni films deposited with 400W $NH_3$ plasma at (a) 200 °C 1.2 Torr, (b) 250 °C 0.6Torr, and (c) 250 °C 1.2 Torr, respectively. As the temperature increased, the nitrogen concentration was found to significantly decrease compared with figure 4 (a) and (c). According to the literature study, the C–N bond contained species were easily desorbed on the Ni(111) surface by dehydrogenation, and nitrile (R–CN) formation caused by thermal energy [19,20]. Therefore, the existence of the C–N bond on the Ni (111) surface closely revealed the possibility of nitrogen desorption. As shown in the XPS results, carbon and nitrogen forms in C–N bonding states. Moreover, the structure of the deposited Ni thin film showed FCC Ni (111) and (200) observed by the XRD measurement (see figure 5(a)).

Accordingly, it can be concluded that higher thermal energy from substrate temperatures causes the thermal desorption of nitrogen easily for every ALD cycle and results in a lower nitrogen impurity concentration. In the case of carbon impurity, the concentration of carbon was dramatically decreased as the process pressure increased as shown in figure 4 (b) and (c). It is well known that $NH_3$ plasma is an effective reducing agent for a metal–organic precursor. Also, $NH_x$ radicals included in $NH_3$ plasma can assist the reduction of metal–organic species during the ALD process [10]. Therefore, it can be concluded that the increment of $NH_x$ radicals in $NH_3$ plasma at 1.2 Torr contributes to more effective reduction of a precursor during the process and causes lower impurity concentrations. However, at 1.8 Torr, it appears that proper reduction cannot occur because the mean free path (MPF) of $NH_3$ plasma was shortened. Accordingly, it can be interpreted that carbon concentration was increased at 1.8 Torr.

As–deposited Ni films were annealed by RTA for 1 minute to form a silicide at an annealing temperature range of 500 to 900 °C. The phase transformation of the ALD Ni film was compared with PVD Ni film. Figure 5 shows the XRD spectra of the as–deposited Ni thin film and nickel silicides after the annealing process using Ni film by (a) the ALD method at 250 °C, 1.2 Torr (denoted by ALD $NiSi_x$) and (b) the PVD method using e–beam evaporation (denoted by PVD $NiSi_x$). The diffraction patterns were observed at the glancing-incident mode from the 2θ range of 25 to 60°. As previously mentioned, the structure of the as–deposited Ni film by RPALD was a face–centered cubic (FCC) structure with a strong peak of (111) corresponding to 44.5° and another (200) crystal plane corresponding to 51.8° (JCPDS No. 04–0850) as presented in figure 5 (a). After annealing for 1 minute to form a silicide phase, low resistive NiSi was formed from 500 °C annealing and its phase was maintained up to 700 °C in both the ALD and PVD $NiSi_x$ samples. The $NiSi_2$ related peaks started to appear at an annealing temperature of over 800 °C. A low resistive NiSi phase still existed in the ALD NiSi$_x$ whereas PVD NiSi$_x$ had only a high resistive NiSi$_2$ phase at this high annealing temperature. The NiSi and NiSi$_2$ phases coexisted in the ALD NiSi$_x$ film after the 800 °C annealing process. However, in the PVD NiSi$_x$, all NiSi was transformed to a high resistive NiSi$_2$ phase with the same condition.

To compare the sheet resistance of the silicide films, FPP measurement was performed. Figure 6 (a) shows the change of sheet resistance by annealing temperature compared with ALD and PVD NiSi$_x$. As shown in the XRD data, both samples formed NiSi at the annealing temperature of 500 to 700 °C and showed around 3–7 Ω/sq.in this temperature range. But, PVD NiSi$_x$ showed relatively rapid increases of sheet resistance between 700 and 800 °C while ALD NiSi$_x$ showed only a slightly change. This difference is thought to be due to the influence of nitrogen and carbon in the as-deposited ALD Ni film. It had been reported that a small amount of nitrogen (under 10 at.%) in the as-deposited Ni thin film contributed to the enhancement of the thermal stability of NiSi [21]. Also, carbons in the Ni thin film played the role of a self-capping layer during the silicide formation [11]. According to the AES depth profile of ALD NiSi$_x$, the high percent of surface carbon, expressed as the 'self-capping layer', was observed while no impurities were contained inside the NiSi film as shown in figure 6 (b). Hence, a certain difference in the thermal stability between ALD and PVD NiSi was anticipated. For this reason, SEM plane view analysis was conducted for both samples.

The morphological stability of the NiSi surface was investigated by SEM observation in the BSE (back–scattered electron) mode. The BSE images contained compositional information due to the difference of the scattering tendency based on the molecular weight. Figure 7 shows the BSE-SEM plane view images of the NiSi surface annealed from (a) ALD Ni and (b) PVD Ni at 700 °C. In the images, a spotted dark phase represented the exposed Si region caused by thermal degradation [22]. As shown in the images, a lot of the Si exposed region was observed in PVD NiSi. Meanwhile, a relatively less Si exposed region was observed in ALD NiSi. These results demonstrate the enhancement of thermal stability of NiSi film from the ALD method. Furthermore, less $R_S$ (sheet resistance) degradation was observed in ALD NiSi$_x$ after the annealing process. Therefore, it can be believed that although the small amounts of carbon and nitrogen were included in the as-deposited ALD Ni film, it contributed to the enhancement of thermal stability of the NiSi layer.

## IV. CONCLUSION

In this study, low resistive Ni thin film was deposited by remote plasma atomic layer deposition with a low impurity concentration. Ni($^i$Pr-DAD)$_2$ was introduced as a Ni precursor and NH$_3$ plasma was used as a precursor. Impurity contents were investigated based on the process temperature and pressure. It was concluded that depending on the temperature increases, nitrogen impurity could be thermally desorbed on the surface of the Ni thin film. Moreover, carbon impurity was mainly affected by process pressure, related to the reaction with NH$_3$ plasma. Small amounts of carbon and nitrogen in as-deposited Ni were found to have contributed to enhancing the thermal stability of NiSi during the RTA process. NiSi from ALD Ni showed better thermal stability compared with those from PVD Ni. This study is meaningful in that Ni film with low resistivity was found to be deposited by the ALD method and formed in the NiSi phase with a higher thermal stability. Therefore it can be believed that the possibility of introducing NiSi for practical industrial applications is more feasible as a result of this study.

# ACKNOWLEDGEMENT

This study was supported by the research program of Hanyang University and SK Hynix Semiconductor (No. 2012–00000260001)

# List of figures

Fig. 1. (a) The growth rate of RPALD Ni and (b) film resistivity for various substrate temperatures

Fig. 2. The impurity concentrations of Ni thin film as a function of (a) substrate temperature at 1.2 Torr and (b) process pressure at 250 °C

Fig. 3. The XPS spectra of impurities core level corresponding to (a) C1s and (b) N1s levels

Fig. 4. AES depth profiles of Ni films deposited for 200 cycles at (a) 200 °C 1.2 Torr, (b) 250 °C 0.6 Torr and (c) 250 °C 1.2 Torr with 400W $NH_3$ plasma.

Fig. 5. The XRD patterns of as-deposited Ni film and nickel-silicide layer by (a) ALD method and (b) PVD method after RTA for 1 min in vacuum ambient at various annealing temperature.

Fig. 6. (a) Sheet resistance of the nickel-silicide films after annealing process for 1 min with various annealing temperatures. (b) AES depth profiles of ALD NiSi at annealing temperature 600 °C

Fig. 7. BES-SEM images of NiSi surface from (a) ALD Ni and (b) PVD Ni at the annealing temperature of 700 °C

**Figure 1**

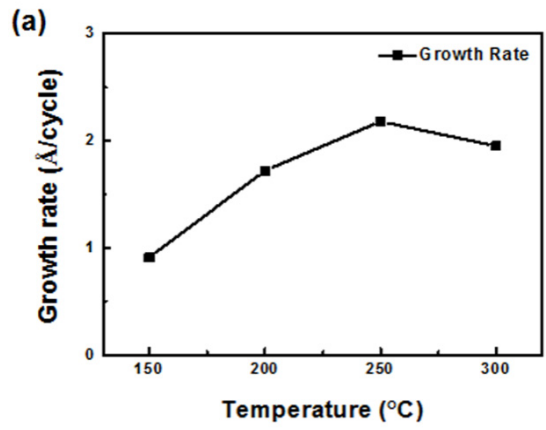

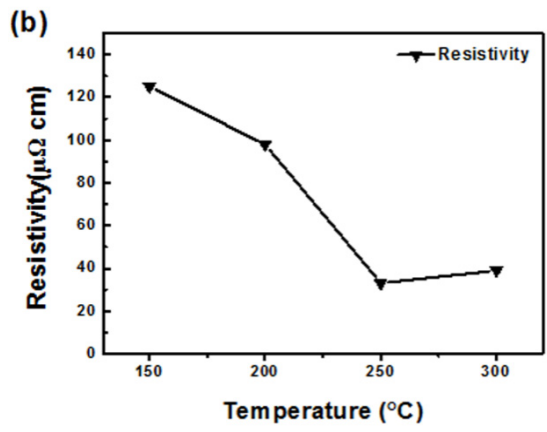

**Figure 2**

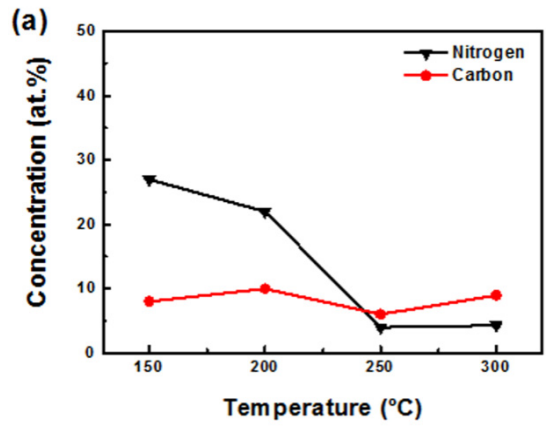

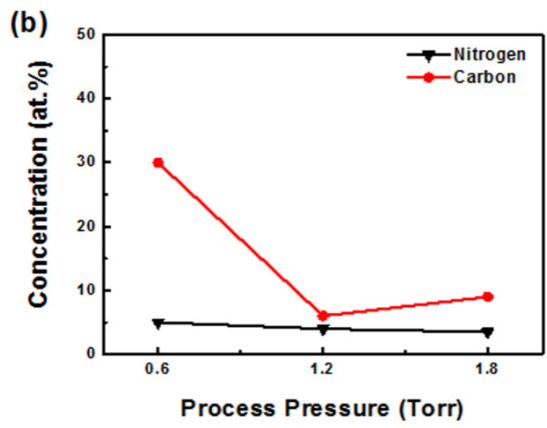

**Figure 3**

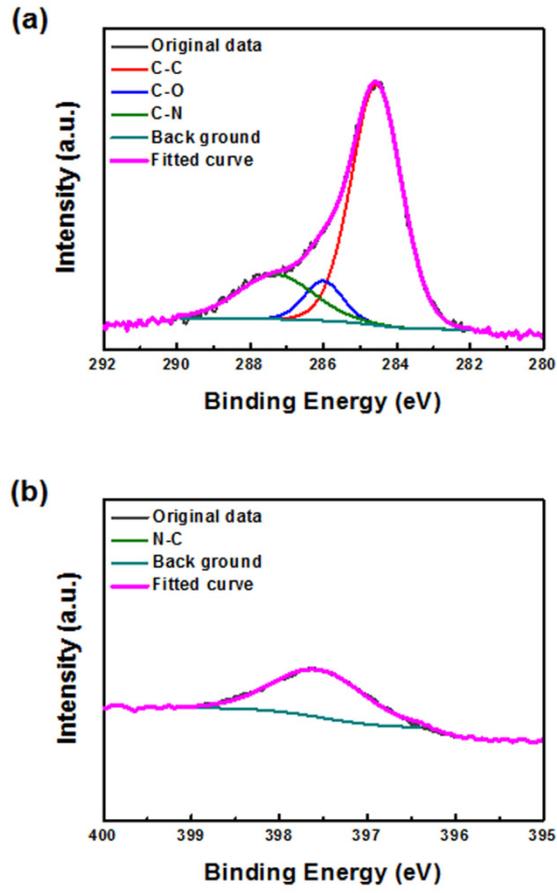

**Figure 4**

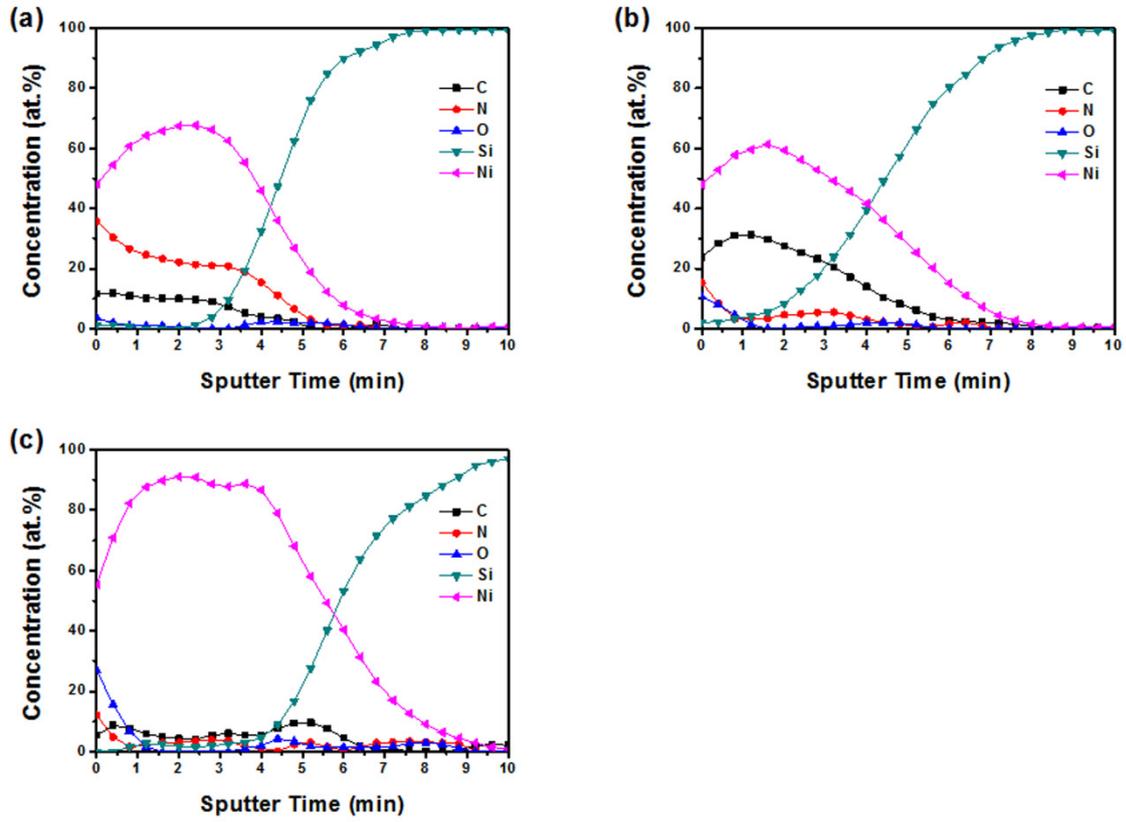

**Figure 5**

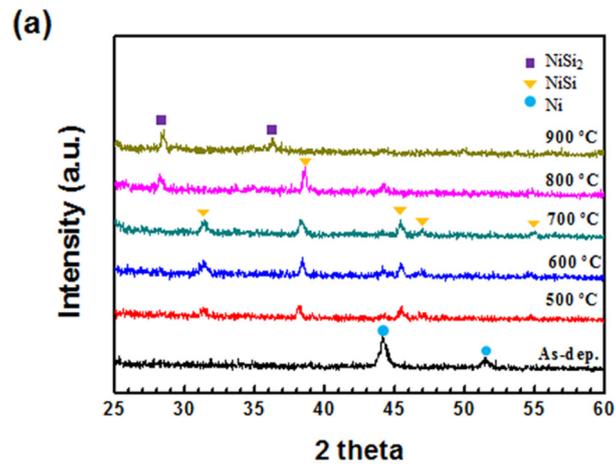

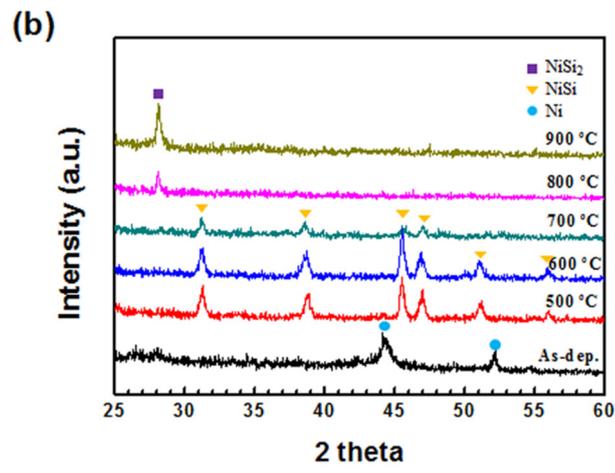

**Figure 6**

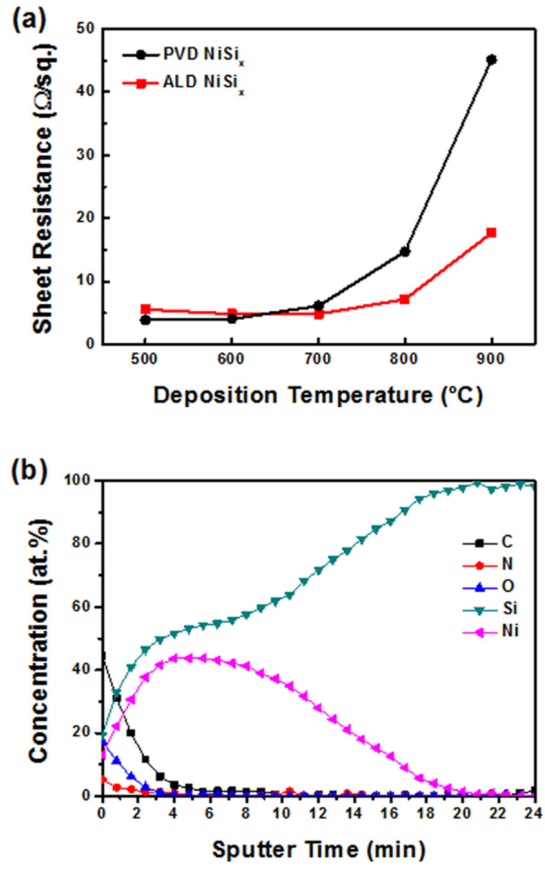

**Figure 7**

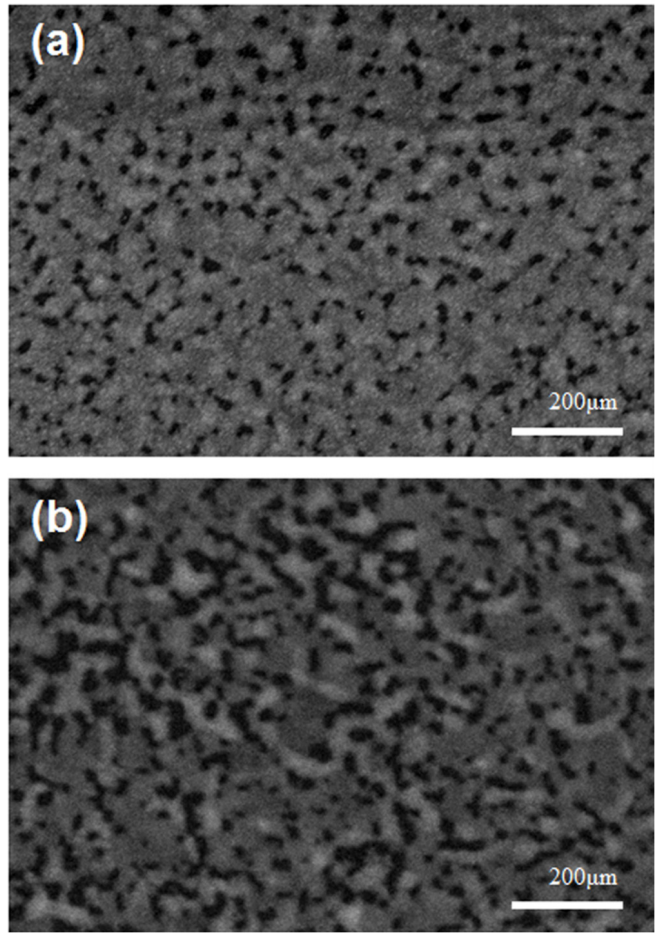